\documentclass[review]{elsarticle}
\usepackage{lineno,hyperref}
\usepackage[english]{babel}
\usepackage[numbers]{natbib}
\usepackage{xcolor}
\usepackage{latexsym,amsmath,amssymb,amsbsy,graphicx,geometry}
\modulolinenumbers[5]

\begin{document}
	
	\begin{frontmatter}
	
\title{Cavity nucleation in single-component homogeneous amorphous solids under negative pressure}

\author[kfu,ufrc]{B.N. Galimzyanov\corref{cor1}}
\cortext[cor1]{Corresponding author}
\ead{bulatgnmail@gmail.com}

\author[kfu,ufrc]{A.V. Mokshin}
\ead{anatolii.mokshin@mail.ru}

\address[kfu]{Kazan Federal University, 420008 Kazan, Russia} 
\address[ufrc]{Udmurt Federal Research Center of the Ural Branch of the Russian Academy of Sciences, 426067 Izhevsk, Russia}

\begin{abstract}
Understanding the cavity formation and cavity growth mechanisms in solids has fundamental and applied importance for the correct determination of their exploitation capabilities and mechanical characteristics. In this work, we present the molecular dynamics simulation results for the process of homogeneous formation of nanosized cavities in a single-component amorphous metallic alloy. To identify cavities of various shapes and sizes, an original method has been developed, which is based on filling cavities by virtual particles (balls) of the same diameter. By means of the mean first-passage time analysis, it was shown that the cavity formation in an amorphous metallic melt is the activation-type process. This process can be described in terms of the classical nucleation theory, which is usually applied to the case of first order phase transitions. Activation energy, critical size and nucleation rate of cavities are calculated, the values of which are comparable with those for the case of crystal nucleation in amorphous systems.
\end{abstract}


\end{frontmatter}

\section{Introduction}

Study of initial stages of the amorphous solids destruction under external deformations is of great importance for the modern materials science as well as for engineering of functional materials~\cite{Guan_Falk_2013,Guo_Jiang_2018,Paul_Dasgupta_2020}. As known from experimental and simulation studies, the process of cavity nucleation can proceed according to homogeneous and heterogeneous scenarios. In the case of the homogeneous scenario, it means such a situation when a fracture center can occur with the same probability in any part of material. Due to the presence of foreign impurities in the samples and sample preparation defects, which serve as foci for the cavity formation in heterogeneous scenario of nucleation, this scenario can be studied experimentally  ~\cite{Lewandowski_2013,Noell_Deka_2022,Das_Elmukashfi_2022}. However, certain difficulties arise in the study of the homogeneous scenario by experimental methods~\cite{Sui_Yu_2022}. On the other hand, both scenarios of the cavity nucleation can be studied using molecular dynamics simulations.

To estimate the cavity nucleation characteristics in solids, the classical nucleation theory developed by Volmer and Weber~\cite{Volmer_Weber_1926} for the case of crystal nucleation and droplet nucleation in condensed matter was applied in 1975 by Raj and Ashby~\cite{Raj_Ashby_1975}. According to their cavity nucleation model, the total free energy $\Delta G$ of the system is determined by the following equation:
\begin{equation}\label{eq_free_energy_cnt}
\Delta G(R) = G_{c}(R)-G_{s}(R)-G_{f}(R).
\end{equation}
Eq.~(\ref{eq_free_energy_cnt}) contains three contributions~\cite{Zhang_2010}: (i) the positive surface energy of a cavity $G_{c}$; (ii) the negative energy $G_{s}$ of a bulk part of the system replaced by a cavity; (iii) the negative relaxation energy $G_{f}(R)$ of elastic deformations caused by a formed cavity. Here, the parameter $R$ is the reaction coordinate characterizing the cavity size. As a rule, the average radius of a cavity is taken as the reaction coordinate assuming that the cavity shape is close to spherical~\cite{Meixner_Ahmadi_2022}. In this case, the basic equations of the classical theory of nucleation designed, for example, to calculate the free energy and the nucleation rate, can be adapted to the case of cavity nucleation without any significant changes~\cite{Meixner_Ahmadi_2022}. Nevertheless, the results of recent studies show that the classical nucleation theory can have limited applicability when the size and the distribution of cavities within the material are weakly dependent on its thermodynamic state~\cite{Sui_Yu_2022,Kassnera_Hayes_2003,Hu_Liu_2022}. Such conditions can be realized, for example, at testing metals for creep~\cite{Hu_Xuan_2021}. Also, for the correct description of the early stage of the cavity formation in the framework of the classical theory, it is great importance to identify correctly stable ``viable'' cavities and to determine the dynamics of changes in their size.

The cavity identification procedure is non-trivial task due to the absence of atoms or molecules recognizing voids within the system. Therefore, the ordinary known cluster analysis criteria and methods used to identify emerging crystalline and/or liquid structures, such as the Voronoi polyhedra method~\cite{Klein_1989}, the Delaunay triangulation method~\cite{Lee_1980}, the method of bond orientational order analysis~\cite{Steinhardt_Nelson_1983,Mickel_Kapfer_2013}, the Stillinger's criterion~\cite{Stillinger_1963,Mokshin_Galimzyanov_2012}, are not applicable in the case of analysis of the cavity formation. To identify the cavities, we can use algorithms, which allows us to reconstruct the geometric shape and to do measurements of the area, volume and curvature gradients of the cavity surface. One of such the algorithms was proposed by Stukowski for 3D-reconstruction of cavities in solids and liquids~\cite{Stukowski_2014}. According to this algorithm, the bulk system is scanned by a virtual spherical probe with the radius $R_{\alpha}$ (here, $R_{\alpha}$ is the position of the main maximum in the radial distribution function). As a result, we can find all the cavities and all the atoms that form the surface of these cavities. Despite the universality and high accuracy, this algorithm does not track the growth of individual cavities.    

In the present work, we propose an original method for detecting cavities inside solids. The method is based on the idea of filling cavities by virtual particles of the same size. Individual cavities are identified and the main parameters of the cavity nucleation such as the activation energy, the critical cavity size, the nucleation waiting time and the nucleation rate are determined on the basis of information about the number of such virtual particles and known coordinates of these particles. In this study, we consider the homogeneous cavity formation in the amorphous metal system at comprehensive expansion. We find that the cavity formation is the process of an activation type and this process can be characterized by means of the mean first-passage time method (MFPT). The applicability of some expressions of the classical nucleation theory to calculation of the activation energy of the critically-sized cavities will also be demonstrated.

\section{Parameters of the system and applied methods}

\subsection{Computational details}

Single-component homogeneous amorphous materials belongs to a family of isotropic solids~\cite{Guan_Falk_2013,Sun_Wang_2015}. The isotropy of such materials and the homogeneity of the amorphous structure minimize the emergence of defects. There are no more obvious preferred areas in amorphous materials, where the emergence of a cavity is more probably. In addition, the growth of a cavity occurs in any direction in a same manner. Therefore, the cavity formation process in these systems occurs according to the homogeneous scenario.

In the present work, we consider the cavity formation in the model amorphous system, where the interatomic interaction is given by the short-range isotropic Dzugutov potential~\cite{Dzugutov_1992,Roth_2000,Galimzyanov_Yarullin_2019}. This potential mimics the well-known features of ion-ion interaction influenced by the electron screening effects as well as the Friedel oscillations in the metals such as Fe, Ni, Cu, Zr, Nb, Pb etc. The system consists $16\,384$ atoms and has the linear size $26.8\,\sigma$, which in the case of metals corresponds to $\sim7.2$~nm (here, $\sigma$ is the effective atom diameter). We apply the method of classical molecular dynamics simulation, where the temperature and the pressure of the system are controlled by thermostat and barostat according to the Nose-Hoover scheme. Samples of the system were prepared at the temperatures $T=0.1\,\epsilon/k_{B}$ and $T=0.2\,\epsilon/k_{B}$ corresponding to deep supercooling levels (here, $\epsilon$ is the energy unit, $k_{B}$ is the Boltzmann constant). Such low temperatures were chosen to obtain a stable amorphous structure without crystalline inclusions. Amorphous samples were prepared through rapid cooling of an equilibrium liquid at the fixed pressure $p=7\,\epsilon/\sigma^3$. The melting temperature of the system at this pressure is  $T_{m}=1.4\,\epsilon/k_{B}$~\cite{Roth_2005}. The temperature of the initial liquid sample is $T=2.5\,\epsilon/k_{B}$. The applied cooling rate is $1\times10^{11}$~K/s. At this cooling rate, the glass transition temperature of the system is $T_{g}=0.6\,\epsilon/k_{B}$. 
\begin{figure}[ht!]
	\centering
	\includegraphics[width=0.6\linewidth]{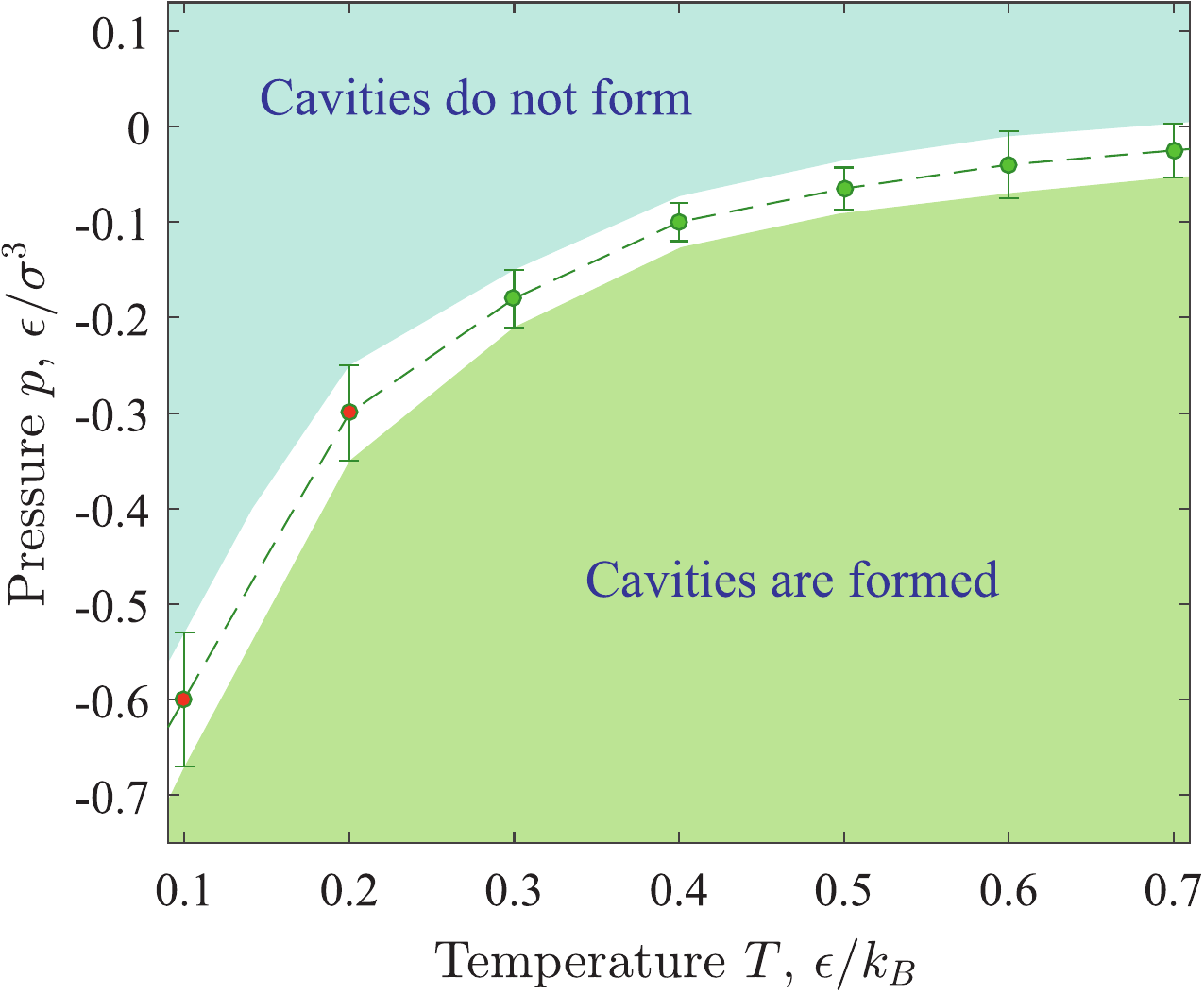}
	\caption{Temperature dependence of the lowest external pressure required to initiate the cavity formation process in Dzugutov system on simulation time scales.}
	\label{fig_1}
\end{figure}  

The simplest way to initialize the cavity nucleation process is the uniaxial tension of the system along some chosen direction. In this case, the cavities will form and grow with increasing deformation of the system. In the present work, the comprehensive stretching is applied to realize the scenario of homogeneous cavity nucleation. According to this method, a constant external negative pressure impacts on the amorphous solid. This leads to emergence of the internal stress that destroys the system. At the same time, the isotropy of the system is kept, which makes the cavity formation equally probable over all the volume of the system. 

For the considered system, negative pressures are given by the Nose-Hoover barostat with the optimal value of the dumping parameter $1000\Delta t$ ($\Delta t=0.005\,\tau$ is the simulation time step)~\cite{Tuckerman_Alejandre_2006}. Before applying negative pressures, the energy of the initial amorphous samples is minimized during the time $2\,\tau$. Additional calculations are carried out to determine the lowest negative pressure. Namely, calculations are performed at various negative pressures during the time $5\,\tau$, where the pressure is changed with increment $0.05\,\epsilon/\sigma^3$ in the range from $p=0$ to $\,\epsilon/\sigma^3$. The pressure at which the formation of cavities is initiated at times up to $5\,\tau$ is considered as the lowest. Calculations are performed in the Lammps package~\cite{Shinoda_Shiga_2004}. We find that the breaking stress is realized for two different conditions: first, when the constant external pressure is $p=-0.6\,\epsilon/\sigma^{3}$ and the temperature is $T=0.1\,\epsilon/k_{B}$; and, second, when the pressure is $p=-0.3\,\epsilon/\sigma^{3}$ and the temperature is $T=0.2\,\epsilon/k_{B}$. As can be seen from Figure~\ref{fig_1}, the larger the system temperature, the easier the system is destroyed and the less negative pressure is required. The taken pressure values at the considered temperatures are the lowest required to initiate the cavity formation process on the time scales available for molecular dynamics studies.

\subsection{Identification of cavities}

We propose the original method, which makes it possible to identify the emerging cavities and their growth dynamics. The proposed method is based on an original procedure for filling cavities by the virtual particles. Obviously, the number of the particles changes, when the cavity evolves with time. The realization of this method takes place in three stages [see Figure~\ref{fig_2}]:
\begin{figure}[ht!]
	\centering
	\includegraphics[width=1.0\linewidth]{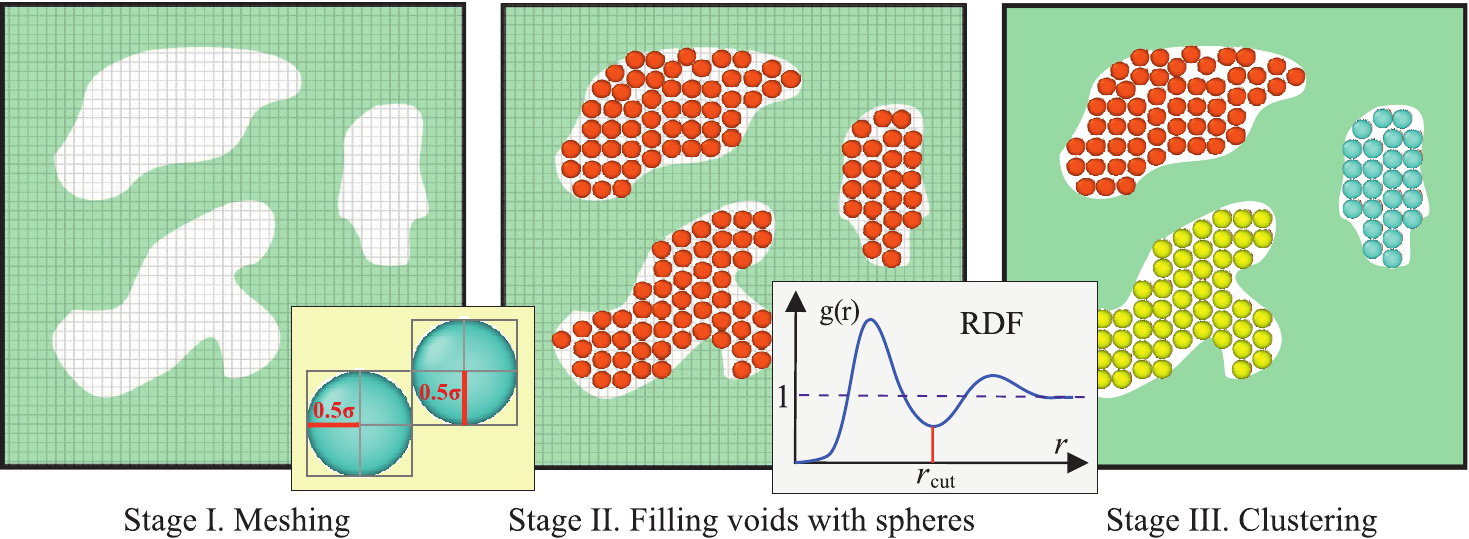}
	\caption{Main stages of the method of filling cavities by virtual particles (balls).}
	\label{fig_2}
\end{figure}  

\noindent 1. It is constructed the three-dimensional grid, which divides the simulation cell into the cubic segments with the equal edge length $\sigma/2$. In the case of the considered system consisting of $16\,384$ atoms, the number of such segments is $154\,000$. All the coordinates of the grid nodes are known [see stage I in Figure~\ref{fig_2}].

\noindent 2. In each node of this grid, a virtual particle of the diameter $\sigma$ is placed. If this particle does not cross with the real atoms of the system and with other virtual particles, then the position of this virtual particle is fixed at the selected grid node. Otherwise, the virtual particle is removed. This procedure continues until all the cavities are filled by the virtual particles [see stage II in Figure~\ref{fig_2}].

\noindent 3. The individual cavities are identified and taken into account at the last stage [see stage III in Figure~\ref{fig_2}]. The number of the virtual particles $\tilde{n}$ located inside each the cavity at different time is determined. As a result, identification of the individual cavities, determination of their sizes and shapes turns into a trivial task. 

In the present study, we apply the original clustering algorithm that allows one to identify individual cavities in the system. The input parameters of this algorithm are the coordinates of the virtual particles and its unique identification numbers. Based on these coordinates and identification numbers, the clustering algorithm determines the nearest environment of each particle. Information about neighboring particles makes it possible to determine the particles belonging to different clusters. As a result, individual clusters of virtual particles are identified and the sizes of these clusters are calculated. Each cluster is assigned a unique identification number to track changes in its size over time. This clustering method is easy to implement and it has a high accuracy in relation to the analysis of molecular dynamics simulation data.

\begin{figure}[ht!]
	\centering
	\includegraphics[width=1.0\linewidth]{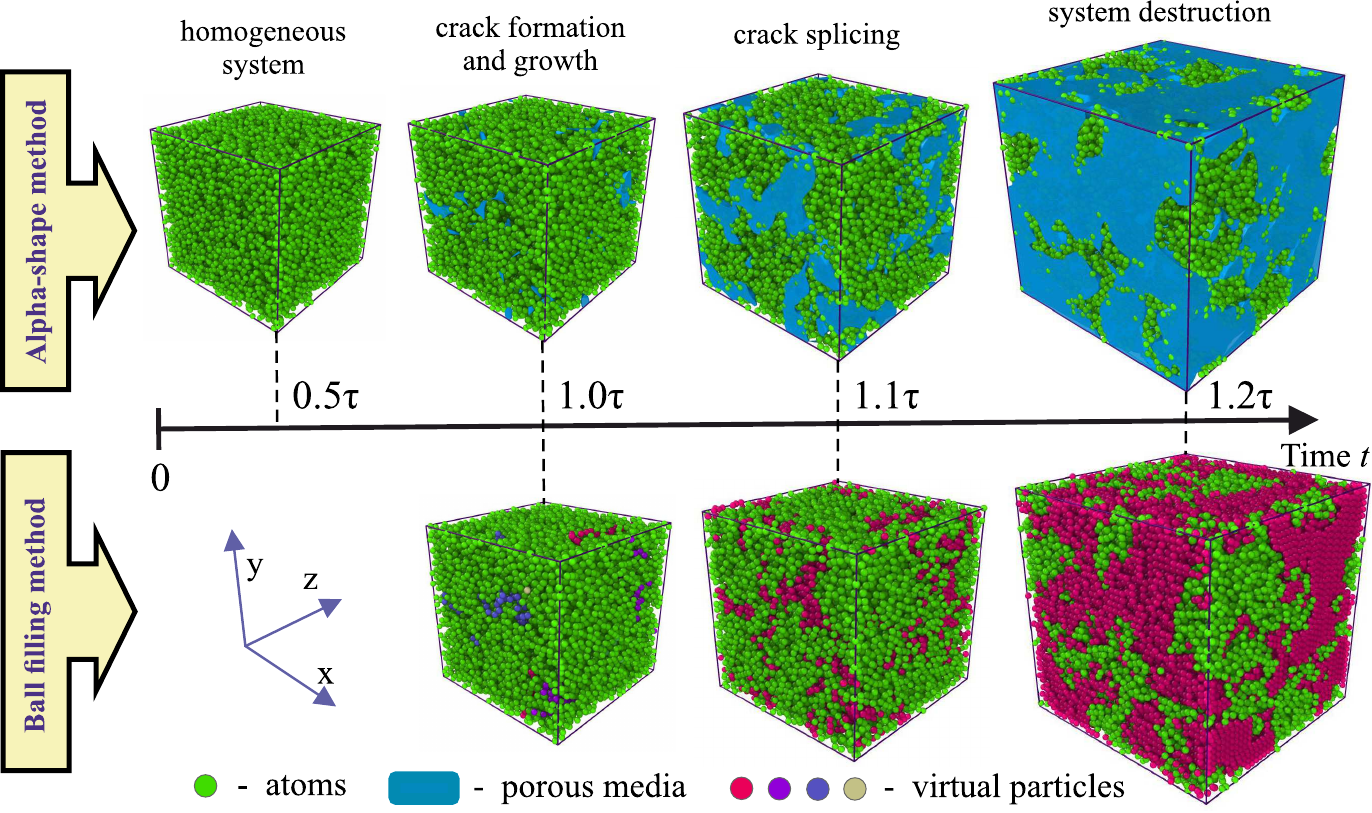}
	\caption{Snapshots of the amorphous system at the comprehensive tension under the negative pressure $p=-0.6\,\epsilon/\sigma^{3}$ at various time moments $t$. The results obtained by filling the cavities by the virtual particles are compared with the results of the Stukowski's method obtained using the OVITO visualization program~\cite{Stukowski_2014}. Both methods correctly identify the location of the cavities. The system atoms are marked in green. The blue area shows the location of the cavities identified by the Stukowski's method (upper snapshots). The result of the proposed method is shown by the virtual particles of different colors (lower snapshots).}
	\label{fig_3}
\end{figure}

As seen from Figure~\ref{fig_3}, the proposed method and the Stukowski's cavity identification method~\cite{Stukowski_2014} produce the same results and correctly display all the emerging cavities inside the considered system. Note that the Stukowski's method allows one to determine the fraction of voids in the system and this method does not identify the individual cavities. In comparison with the Stukowski's method, the filling virtual balls method proposed in the present study allows one to determine individual cavities and to track the change in the size of each of them. The alpha-shape method proposed by Edelsbrunner and M\"{u}cke~\cite{Edelsbrunner_1994} underlies Stukowski's cavity algorithm, in which a probe sphere with the radius $R_\alpha$ is applied to detect voids inside a solid. The probe sphere scans a solid and detects regions of space without any atoms. The cavity texture depends on the radius $R_\alpha$: the smaller the radius, the small cavities are better identified. The optimal value of the radius $R_\alpha$ is equal to the distance between the nearest neighbors of atoms. The atoms that form the surface of a solid are also detected by the probe sphere. Delaunay triangulation of the input point set (i.e., positions of the surface atoms) is applied to construct the surface mesh of a solid. The resulting triangles are connected and form a closed surface mesh that separates the solid from the empty regions. 

\section{Cavity formation and coordination analysis}
	
An important condition for the formation of the ``viable'' cavity is the reaching the certain critical size $\widetilde{n}_c$, which determines the number of the virtual particles located inside this cavity. If the size of the cavity is less than $\widetilde{n}_c$, then the attraction forces between the separated atoms can lead to cavity collapse. In the case of formation of the cavity larger than $\widetilde{n}_c$, the interatomic attraction forces are not sufficient to hold ``the surface atoms''. This leads to further increase of the cavity size. Thus, the cavity formation can be compared with the process of homogeneous crystal nucleation in supercooled liquids and amorphous solids, where crystallization is also initiated after the appearance of critically-sized nucleus~\cite{Kelton_Greer_2010,Mokshin_Galimzyanov_2017,Malek_Morrow_2015,Galimzyanov_Yarullin_2018}. 

\begin{figure}[ht!]
	\centering
	\includegraphics[width=1.0\linewidth]{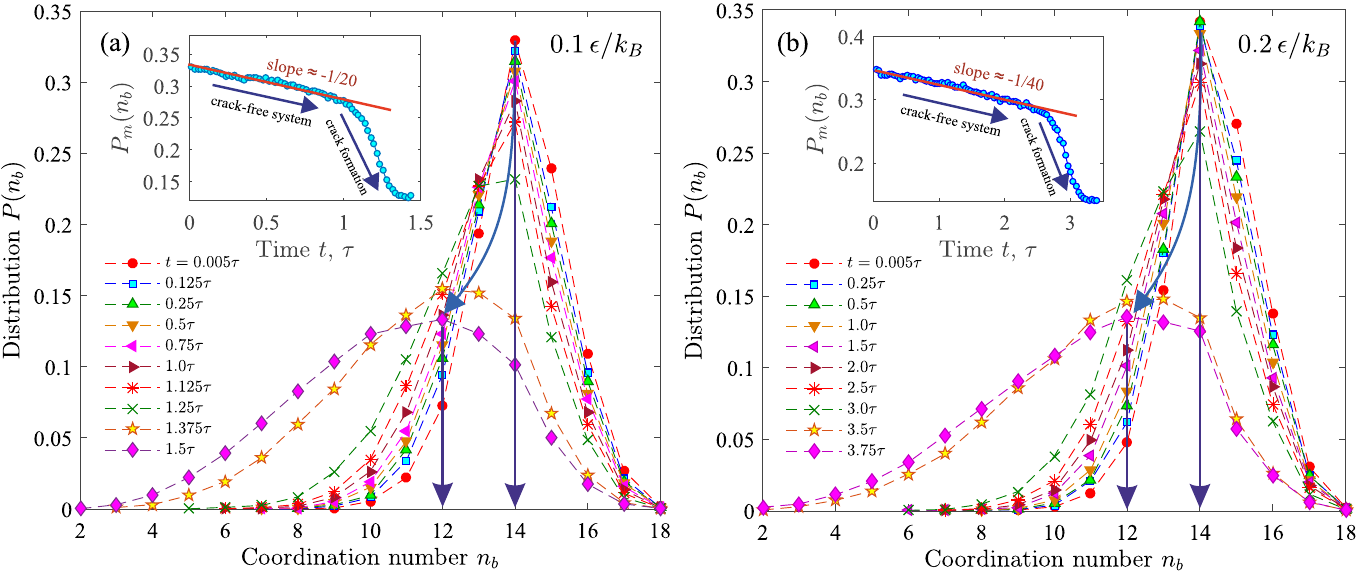}
	\caption{Distribution function $P(n_b)$ over the coordination number $n_b$ calculated at different time moments and at two different temperatures: (a) $T=0.1\,\epsilon/k_{B}$; (b) $T=0.2\,\epsilon/k_{B}$. Insets: correspondence between the maximum of the distribution $P_{m}(n_{b})$ and the time $t$, where the arrows show different regimes.}
	\label{fig_4}
\end{figure}

It follows from the above that the intensive cavity formation must be accompanied by significant change of the number of the nearest neighbors $n_{b}$ for atoms. Figure~\ref{fig_4} shows the distribution function $P(n_b)$ of the coordination number $n_b$ calculated at two different temperatures and at different time moments. Panels (a) and (b) of Figure~\ref{fig_4} show that the shape of these distributions weakly depends on the system temperature. The maximum value of the coordination number $n_{b}$ in the calculated distributions $P(n_b)$ decreases from $14$ atoms to $12$ atoms. The value $n_{b}=14$ atoms corresponds to the homogeneous system without cavities, whereas the value $n_{b}=12$ atoms corresponds to the system destroyed due to the formation and subsequent coalescence of large cavities. In this case, the coordination number starts to decrease only after the waiting time $\widetilde{\tau}_c$, which is necessary to formation of the cavities of the critical size $\widetilde{n}_{c}$.

As can be seen from the insets to Figure~\ref{fig_4}, the time dependence of the maximum of the distribution function, $P_m(n_b)$, has two regimes. The first regime characterizes the system without cavities, where the quantity $P_m(n_b)$ decreases linearly with time $t$ and is reproduced by a straight line. The larger the system temperature, the smaller the negative slope of this line due to the thermal expansion of the system. The second regime is accompanied by the rapid decrease in the value of the quantity $P_m(n_b)$ due to the intensive growth of cavities and due to the increase of the number of atoms that form the surface of the cavity.

\section{MFPT-analysis and classical nucleation theory}

\begin{figure}[ht!]
	\centering
	\includegraphics[width=0.6\linewidth]{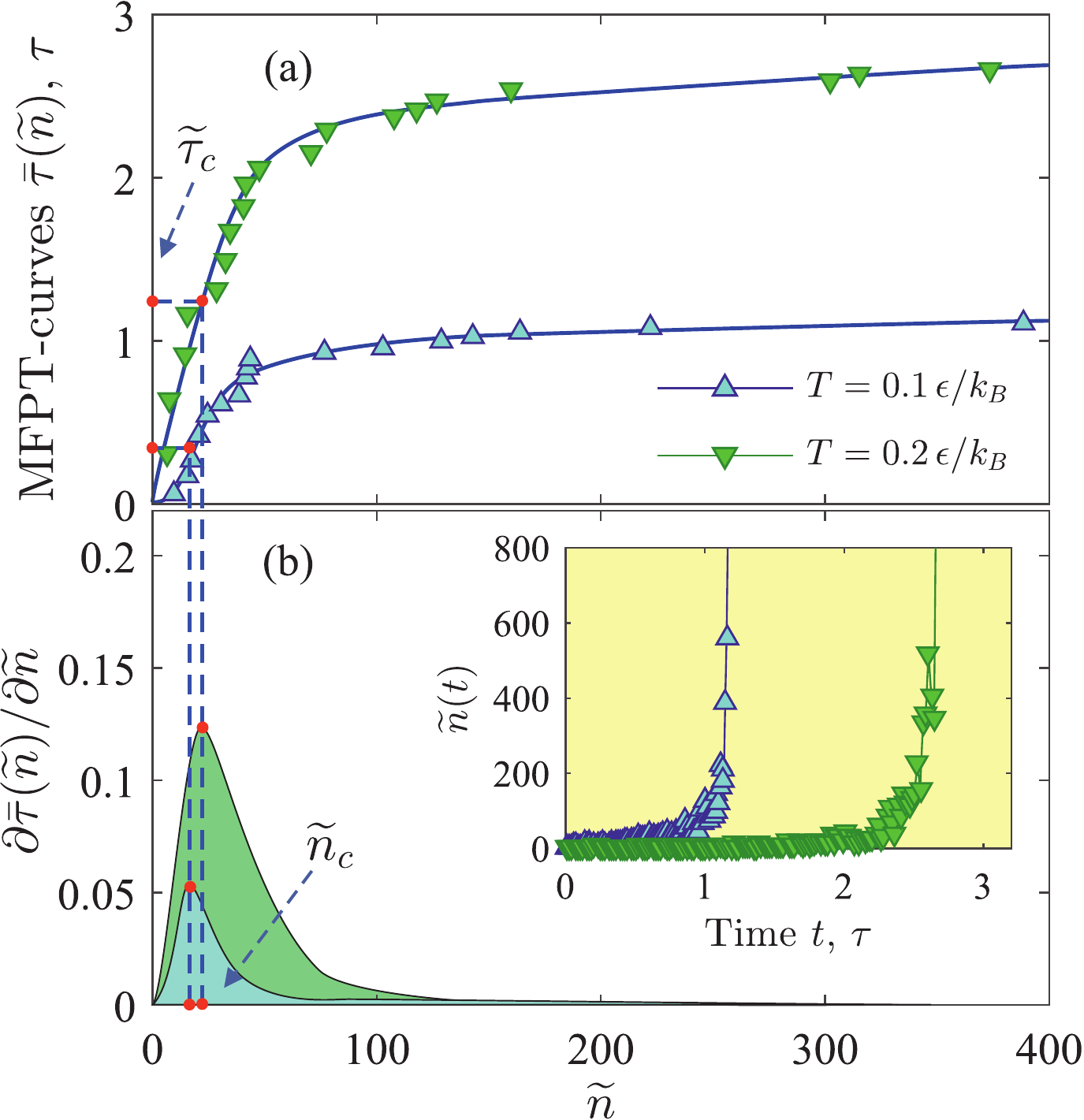}
	\caption{(a) MFPT-curves computed at different temperatures. (b) Derivatives $\partial\bar{\tau}(\widetilde{n})/\partial\widetilde{n}$ of the MFPT-curves. The arrows show the inflection points and positions of extrema in the obtained dependences $\bar{\tau}(\widetilde{n})$ and $\partial\bar{\tau}(\widetilde{n})/\partial\widetilde{n}$ corresponding to the critical size $\widetilde{n}_c$ and the waiting time $\widetilde{\tau}_c$. Insert: growth trajectories $\widetilde{n}(t)$ of the largest cavity at various temperatures.}
	\label{fig_5}
\end{figure}

Estimation of the average value of the critical size $\widetilde{n}_{c}$ and the waiting time $\widetilde{\tau}_{c}$ for the first critically-sized cavity was carried out through observing the change in the size of the largest cavity. Based on the results of $20$ independent  molecular dynamics iterations, the most probable (averaged) growth trajectories $\widetilde{n}(t)$ of the large cavity were determined at various temperatures [see inset in Figure~\ref{fig_5}(b)]. The parameter $\widetilde{n}(t)$ characterizes the time dependence of the number of virtual particles placed inside the cavity. At the considered thermodynamic conditions, the time dependence of the size $\widetilde{n}$ is typical for activation-type processes: the cavity size fluctuates around zero for some time and starts to grow only after formation of a stable cavity.

The MFPT-curves $\bar{\tau}(\widetilde{n})$~\cite{Mayo_Perkins_2011,Nicholson_Rutledge_2016} were computed using the method of inverted averaging of growth trajectories described in detail in Ref.~\cite{Mokshin_Galimzyanov_2012,Mokshin_Galimzyanov_2014}:
\begin{equation}\label{eq_mfpt}
\bar{\tau}(\widetilde{n})=\frac{1}{M}\sum_{\alpha=1}^{M}t_{\alpha}(\widetilde{n}).
\end{equation}
Here, $t_{\alpha}(\widetilde{n})$ is the inverted growth trajectory of the largest cavity in $\alpha$-th molecular dynamics iteration and $M=20$. Figure~\ref{fig_5}(a) shows that the MFPT-curves have the sigmoidal shape, where the position of the inflection point determines the average critical size $\widetilde{n}_{c}$ at the waiting time $\widetilde{\tau}_{c}\equiv\bar{\tau}(\widetilde{n}=\widetilde{n}_{c})$. The exact values of the parameters $\widetilde{n}_{c}$ and $\widetilde{\tau}_{c}$ were determined by differentiate these MFPT-curves~\cite{Manuel_2006}. To properly derive the derivative from the MFPT data presented on Figure~\ref{fig_5}(a), we approximate this data by the $8$th degree polynomial function
\begin{equation}\label{eq_fit_func}
\bar{\tau}(\widetilde{n})=a_{0}+\sum_{i=1}^{8}a_{i}\widetilde{n}^{i}.
\end{equation}
Here, the values of the parameters $a_i$ are determined during the fitting process. The value of the parameter $a_0$ is zero. It should be noted that the $8$th degree polynomial is sufficient to correctly reproduce the inflection region in obtained MFPT-curves and to determine the cavity nucleation characteristics. 

As can be seen from Figure~\ref{fig_5}(b), the quantity $\partial\bar{\tau}(\widetilde{n})/\partial\widetilde{n}$ contains a single extremum (maximum), whose position determines the value of the critical size $\widetilde{n}_c$. We find that the critical size is $\widetilde{n}_{c}=(25\pm2)$ virtual particles for the case of the system with the temperature $T=0.1\,\epsilon/k_{B}$ and $\widetilde{n}_{c}=(30\pm3)$ virtual particles for system with the temperature $T=0.2\,\epsilon/k_{B}$. At these temperatures, the values of the waiting time for the critically-sized cavities are $\widetilde{\tau}_{c}=(0.37\pm0.02)\,\tau$ and $\widetilde{\tau}_{c}=(1.15\pm0.05)\,\tau$ respectively. The increase of the critical size $\widetilde{n}_{c}$ and the waiting time $\widetilde{\tau}_{c}$ with temperature is because of formation of the larger cavities becomes energetically favorable at the high temperatures: the faster the thermal motion of atoms, the larger must be the critical size $\widetilde{n}_{c}$ so that the cavity cannot collapse.

The activation energy $\Delta\widetilde{G}$ for the critically-sized cavity was estimated based on the MFPT data. Here, we assume that the shape of the cavities nucleating in the isotropic amorphous system is close to spherical at the considered thermodynamic conditions, and the activation barrier $\Delta\widetilde{G}_{c}/k_{B}T$ is symmetrical. Then, the value of the quantity $\Delta\widetilde{G}$ can be estimated by the following expression (see section ``Appendix''): 
\begin{equation}\label{eq_energybarrier}
\frac{\Delta\widetilde{G}_{c}}{k_{B}T}=3\pi(\widetilde{n}_{c}\widetilde{Z})^{2},
\end{equation}  
known in the classical nucleation theory~\cite{Kelton_Greer_2010,Huitema_2000}. It should be noted that, the quantity $\Delta\widetilde{G}_{c}$ in Eq.~(\ref{eq_energybarrier}) will characterize the energy required to break of the interatomic attraction forces and to form the cavity with the critical size $\widetilde{n}_{c}$. The parameter $\widetilde{Z}$ known as the Zeldovich factor determines the curvature of the activation barrier. In Ref.~\cite{Wedekind_Strey_2007}, it was shown that the curvature around the top of the energy barrier $\Delta G(n)$ is related to the shape of the MFPT-curve $\bar{\tau}(\widetilde{n})$. Namely, the MFPT-curve has the inflection, the slope of which determines the curvature of the nucleation barrier $\Delta G(n)$: the greater the slope, the greater the curvature $\Delta G(n)$. Then the first derivative of the MFPT-curve $\partial\bar{\tau}(\widetilde{n})/\partial\widetilde{n}$ will characterize the curvature of the nucleation barrier and will be related to the Zeldovich factor $\widetilde{Z}$ through the expression~\cite{Mokshin_Galimzyanov_2012,Cai_Wu_2014}
\begin{equation}\label{eq_Z}
\widetilde{Z}=\frac{1}{\tau_{s}}\frac{\partial\bar{\tau}(\widetilde{n})}{\partial\widetilde{n}}\Bigg|_{\widetilde{n}=\widetilde{n}_{c}}.
\end{equation} 
Here, the parameter $\tau_{s}$ is the nucleation time scale, which can be defined from the MFPT-curve as $\tau_s\simeq2\widetilde{\tau_{c}}$. In this case, the parameter 
$\tau_s$ will be related with the nucleation rate $\widetilde{J}_{st}$ of cavities via the expression
\begin{equation}\label{eq_Jst}
\widetilde{J}_{st}=\frac{1}{\tau_{s}V}\simeq\frac{1}{2\widetilde{\tau_{c}}V},
\end{equation}
where $V$ is the volume of the system. Eq.~(\ref{eq_Jst}) makes it possible to determine the nucleation rate of the first critically-sized nucleus. However, in the case of a system whose energy has been minimized, this rate $\widetilde{J}_{st}$ corresponds to the steady-state nucleation rate. 

The obtained results reveal that the nucleation time scales are $\tau_s=(0.74\pm0.02)\,\tau$ at the temperature $T=0.1\,\epsilon/k_{B}$ and $\tau_s=(2.3\pm0.05)\,\tau$ at the temperature $T=0.2\,\epsilon/k_{B}$. The maxima in the derivatives of the MFPT-curves are $\partial\bar{\tau}(\widetilde{n})/\partial\widetilde{n}|_{\widetilde{n}=\widetilde{n}_ {c}}\simeq0.053\,\tau$ and $\partial\bar{\tau}(\widetilde{n})/\partial\widetilde{n}|_{\widetilde{n}=\widetilde{n}_{ c}}\simeq0.126\,\tau$ respectively. Then, at the considered temperatures, Eq.~(\ref{eq_Z}) gives the values $\widetilde{Z}=(0.0716\pm0.004)$ and $\widetilde{Z}=(0.0547\pm0.007)$ respectively. The parameter $\widetilde{Z}$ takes the values in the interval $[0.01;\,1]$ that is reasonable in the framework of the classical nucleation theory~\cite{Kashchiev_2000}. Then, from Eq.~(\ref{eq_energybarrier}) we find the following values for the activation barrier: $\Delta\widetilde{G}_{c}/k_{B}T=(30.2\pm2.1)$ (at $T=0.1\,\epsilon/k_{B}$) and $\Delta\widetilde{G}_{c}/k_{B}T=(34.5\pm2.5)$ (at $T=0.2\,\epsilon/k_{B}$). At the considered temperatures, the energies required to form of the critically-sized cavity are $\Delta\widetilde{G}_{c}\simeq3.02\,\epsilon$ and $\Delta\widetilde{G}_{c}\simeq6.9\,\epsilon$. It is noteworthy that the found values of the quantity $\Delta\widetilde{G}_{c}$ are approximately two times larger than the activation energies calculated for the case of crystal nucleation in the amorphous Dzugutov system. For example, to the formation of critically-sized crystalline nuclei in the amorphous Dzugutov system, it is required energies equal to $\Delta G_{c}\simeq1.5\,\epsilon$ and $\Delta G_{c}\simeq2.9\,\epsilon$ at the temperatures $T=0.1\,\epsilon/k_{B}$ and $T=0.2\,\epsilon/k_{B}$ but at the pressure $p=14\,\epsilon/k_{B}$~\cite{Mokshin_Galimzyanov_2017_RSB}. Moreover, from Eq.~(\ref{eq_Jst}) we find that the steady-state nucleation rate takes the values $\widetilde{J}_{st}\simeq6.61\times10^{-5}\tau^{-1}\sigma^{-3}$ (at $T=0.1\,\epsilon/k_{B}$) and $\widetilde{J}_{st}\simeq2.04\times10^{-5}\tau^{-1}\sigma^{-3}$ (at $T=0.2\,\epsilon/k_{B}$), which are also comparable with the crystal nucleation rate in supercooled liquids and amorphous systems~\cite{Huang_Chen_2018,Mokshin_Barrat_2010,Mokshin_Galimzyanov_2013}.

\section{Conclusions}

Our studies show that the process of homogeneous cavity nucleation in single-component amorphous systems proceeds according to the scenario that is similar to the crystal nucleation in these systems and can be described in terms of the classical nucleation theory~\cite{Raj_Ashby_1975,Kelton_Greer_2010,Kashchiev_2000}. For the case of homogeneous cavity formation in the Dzugutov's amorphous system, it is shown that the formation of the stable cavity is the activation-type process. The realization of the activation-type process is confirmed by the presence of two regimes in the time dependences of the coordination number distributions. The transition between these regimes occurs due to jump-like decrease in the number of nearest neighbors from $14$ to $12$ atoms. By the method of the MFPT-analysis, the critical size and the waiting time for the critically-sized cavity were calculated; and the values of the parameters characterizing the curvature of the activation barrier were also calculated. We have shown that Eqs. (\ref{eq_energybarrier}), (\ref{eq_Z}) and (\ref{eq_Jst}) as applied to the case of cavity formation allow one to determine the activation energy $\Delta\widetilde{G}_{c}$, the Zeldovich factor $\widetilde{Z}$ and the steady-state nucleation rate $\widetilde{J}_{st}$.

To identify emerging cavities, an original method was proposed based on filling cavities by the virtual particles of the same diameter. The proposed method correctly identifies cavities of various shapes and sizes, that is confirmed by good agreement with the results of 3D visualization by means of the Stukowski's method~\cite{Stukowski_2014}. Using the proposed method, the sizes of the cavities in terms of the number of the virtual particles are determined, which made it possible to perform the MFPT-analysis and to interpret the results within the classical nucleation theory. The obtained results characterize the features of cavity formation and do not depend on the compositions of a system. Therefore, these results can be directly extended to the multiple-component amorphous solids and it can be used to develop a unified theory of cavity nucleation and growth in condensed matter with different types of interatomic interaction. Moreover, the results of the present study can be applied to the case of dynamic loading of amorphous solids~\cite{Jiang_Meng_2008,Huang_Ling_2014}. In this case, it should be taken into account that cavity formation at dynamic loading corresponds to the case of structural transformations under non-equilibrium conditions. Therefore, for this case, the theoretical description could be done in the framework of a modified theory, which captures these effects (e.g. effective temperature)~\cite{Mokshin_Galimzyanov_2013}.

\section*{Acknowledgment}
\noindent The work was supported by the Russian Science Foundation (project No. 19-12-00022-P). Review part of the work was done with the support by the Kazan Federal University Strategic Academic Leadership Program (PRIORITY-2030).

\section*{Appendix: Derivation of the expression for the nucleation barrier}

According to the classical nucleation theory~\cite{Kelton_Greer_2010,Sosso_2016}, the free energy $\Delta G$ required for the formation of a new phase nucleus with the radius $r$ is determined by the expression
\begin{equation}\label{eqA_1}
\Delta G(r)=4\pi\sigma_{s}r^{2}-\frac{4\pi}{3}\rho|\Delta\mu|r^{3}.
\end{equation}
Here, $\sigma_{s}$ is the surface tension of a nucleus; $\rho$ is the number density of a nucleus; $|\Delta\mu|$ is the difference between the chemical potentials of the parent and daughter phases. Expression (\ref{eqA_1}) is valid for a spherical nucleus with the averaged radius $r$. Then, using the relation $r=[3n/(4\pi\rho)]^{1/3}$, expression (\ref{eqA_1}) can be presented in terms of the nucleus size $n$: 
\begin{equation}\label{eqA_2}
\Delta G(n)=4\pi\sigma_{s}\left(\frac{3n}{4\pi\rho}\right)^{2/3}-n|\Delta\mu|.
\end{equation}
Here, $n$ is the number of atoms in a nucleus with the radius $r$.

In the case of critically-sized nucleus with the size $n_c$, the presence of a nucleation barrier is determined by the condition~\cite{Clouet_2009,Kalikmanov_2013}
\begin{equation}\label{eqA_3}
\frac{\partial\Delta G(n)}{\partial n}\Bigg|_{n=n_c}=0.
\end{equation}
Then from (\ref{eqA_2}) and (\ref{eqA_3}) the expression for the critical size $n_c$ is obtained as follows~\cite{Sosso_2016}:
\begin{equation}\label{eqA_4}
n_c=\frac{32\pi}{3}\frac{\sigma_{s}^{3}}{\rho^{2}|\Delta\mu|^{3}}.
\end{equation}
Substituting (\ref{eqA_4}) to (\ref{eqA_2}), we obtain the expression for the free energy $\Delta G_{n_c}$ required for the formation of a critically-sized nucleus:
\begin{equation}\label{eqA_5}
\Delta G_{n_c}=\frac{16\pi}{3}\frac{\sigma_{s}^{3}}{(\rho|\Delta\mu|)^{2}}.
\end{equation}
Thus, from (\ref{eqA_4}) and (\ref{eqA_5}) we found the relation
\begin{equation}\label{eqA_6}
\Delta G_{n_c}=\frac{|\Delta\mu|n_c}{2}.
\end{equation}

The flatness of the free energy profile around the critical nucleus size is characterized by the Zeldovich factor $Z$. As known~\cite{Kashchiev_2000}, the factor takes values in the range $Z\in[0.01;\,1]$. This factor is the function of the second derivative of the free energy $\Delta G(n)$ at the critical size $n_c$~\cite{Clouet_2009}:
\begin{equation}\label{eqA_7}
Z=\sqrt{-\frac{1}{2\pi k_{B}T}\frac{\partial^{2}\Delta G(n)}{\partial n^{2}}\Bigg|_{n=n_c}}.
\end{equation}
Here, $T$ is the system temperature. Then from (\ref{eqA_2}) and (\ref{eqA_7}) we find the expression  
\begin{equation}\label{eqA_8}
Z=\sqrt{\frac{(36\pi)^{1/3}}{9\pi k_{B}T}\frac{\sigma_{s}n_c^{-4/3}}{\rho^{2/3}}}.
\end{equation}
Taking into account (\ref{eqA_4}), (\ref{eqA_6}) and (\ref{eqA_8}), the equation for the activation barrier is found as follows:
\begin{equation}\label{eqA_9}
\frac{\Delta G_{n_c}}{k_{B}T}=3\pi(n_c Z)^{2}.
\end{equation}
Thus, considering that for the case of cavity formation we apply $n_c\equiv\widetilde{n}_{c}$, $Z\equiv\widetilde{Z}$ and $\Delta G_{n_c}\equiv\Delta \widetilde{G}_{c}$, expression (\ref{eqA_9}) is identical to (\ref{eq_energybarrier}).

\end{document}